\documentclass[preprint,authoryear,11pt,dvipsnames,svgnames,table]{elsarticle}

\makeatletter

\providecommand\IfPDFManagementActiveT[1]{}

\makeatother

\usepackage{graphicx}
\usepackage{amssymb}
\usepackage{amsthm}
\usepackage{amsmath}
\usepackage{float}
\usepackage{ctable}
\usepackage[margin=1.2in]{geometry}
\usepackage{eurosym}
\usepackage{longtable, booktabs}
\biboptions{sort,round}
\usepackage{multirow}  
\usepackage{mathpazo}
\usepackage{url} 
\usepackage{enumitem}
\usepackage{booktabs}
\usepackage{threeparttable}
\usepackage{caption}
\usepackage{xcolor}
\usepackage{dcolumn}

\definecolor{mybox}{HTML}{3333B2}
\definecolor{mygreen}{HTML}{007F00}
\definecolor{myred}{HTML}{E10000}
\definecolor{myorange}{HTML}{FF8000}

\journal{}

\makeatletter

\newcommand{\sym}[1]{\ifmmode^{#1}\else\(^{#1}\)\fi}

\begin{document}

\begin{frontmatter}

\title{Do Short Exposure and Systematic Risk Exposure Drive Asymmetries in the Disposition Effect?}

\author[c]{Lorenzo Mazzucchelli\corref{cor1}}
\ead{achab.lorenzo@gmail.com}
\author[d]{Marco Zanotti}
\ead{m.zanotti22@campus.unimib.it}
\author[c]{Luca Vincenzo Ballestra}
\ead{luca.ballestra@unibo.it}
\author[c]{Andrea Guizzardi}
\ead{andrea.guizzardi@unibo.it}

\address[c]{University of Bologna, Department of Statistical Sciences, \\Via delle Belle Arti 41, 40126 Bologna, Italy\vspace{6pt}}

\address[d]{University of Milano-Bicocca, Department of Economics, Management and Statistics, \\Piazza dell'Ateneo Nuovo 1, 20126 Milano, Italy}

\cortext[cor1]{Corresponding author}

\begin{abstract}

This study examines the disposition effect in both long and short exposure positions in FTSE MIB-tracking ETFs using a unique dataset of almost 9 million individual transactions. Building on the integrated framing approach, we extend the analysis to explicitly incorporate leverage and long/short exposures, allowing us to assess how portfolio context and systematic risk exposure jointly are associated to investors’ realization behavior. 
Methodologically, we generalize Odean’s canonical Count and Total measures to wide and integrated framing, introduce a novel Value metric that captures the return thresholds required to realize gains versus losses, and implement these measures in \texttt{dispositionEffect}, an open-source R package for large-scale intraday data.
We show that short positions exhibit a weaker disposition effect than long positions under narrow framing, but that this asymmetry reverses in positively performing portfolios under integrated framing. Systematic risk further amplifies these behavioral asymmetries across positions. Overall, our findings demonstrate that the disposition effect is not solely asset-specific, but is critically shaped by the interaction between portfolio context, position type, and systematic risk exposure.  More broadly, the results are consistent with the joint predictions of Prospect Theory and Regret Theory, highlighting the central role of framing in investor decision-making.


\vspace{6pt}
\end{abstract}


\end{frontmatter}

\newpage


\section{Introduction}
\label{sec:intro}

Research on the causes of the disposition effect has evolved significantly over the past few decades. Initially formulated by Shefrin and Statman (1985), this phenomenon describes investors' tendency to sell winning assets too early while holding losing assets too long. 

From a theoretical perspective, it challenges the assumptions of rational investor behavior in traditional finance models and provides insights into the psychological mechanisms driving financial decisions. From a practical standpoint, the disposition effect has important implications for market efficiency, asset pricing, and investor welfare, as it can lead to suboptimal portfolio choices and market anomalies (Jegadeesh and Titman, 1993). 

Despite extensive research since its inception, several aspects of the disposition effect remain inadequately explored, particularly in the context of contemporary trading environments. The emergence of new trading technologies and instruments has significantly increased the adoption of short exposure and leverage practices even among non-professional investors (Barber and Odean, 2002; Geczy et al., 2002). This creates a new landscape where the disposition effect may manifest differently, making it imperative to explore how it relates to short exposure and leverage among individual investors.
At the same time, the increasing availability of high frequency trading records calls for methodological frameworks that can exploit investor level intraday data while remaining comparable with the canonical literature.

We address this challenge by analyzing the disposition effect on a large sample of retail investors trading FTSE MIB-linked leveraged ETFs, including both long and inverse products, observed at intraday frequency. Short exposure in our setting is achieved through inverse leveraged ETFs, which enables us to compare realization decisions across long and inverse positions linked to the same underlying index. 
Our dataset, sourced from an Italian brokerage, includes approximately 4,600 retail investors, more than 19,000 traded assets and almost 9 million transactions over the 2010–2018 period. For the empirical analysis, we focus on transactions in eight FTSE MIB-linked leveraged ETFs, while the broader asset universe is used to reconstruct investors’ portfolio states in real time.
Building on recent advances in wide and integrated framing (Brettschneider et al., 2021; Ballestra et al., 2024), we implement an integrated framing approach that jointly considers stock-level and portfolio-level information and explicitly incorporates the long/short dimension.
This unified setting allows us to study how portfolio context, position type, and systematic risk exposure jointly shape realization decisions, moving beyond asset by asset analyses.

To the best of our knowledge, no previous study has combined investor‑level intraday data, short‑exposure activity, and portfolio framing within a unified empirical framework.
Existing work on short exposure and the disposition effect, such as Von Beschwitz and Massa (2020), relies on aggregate data and does not fully account for portfolio context, nor does it disentangle long and inverse positions within the same investor. 
Similarly, studies on risk and the disposition effect typically focus on idiosyncratic or perceived risk, leaving the role of systematic market exposure only partially identified.

Our integrated framing setup closes this gap by assessing disposition effects for long and short ETFs conditional on the aggregate portfolio being in a gain or loss state. Our approach also introduces a dynamic, investor‑specific time frame for decision‑making, contrasting with the fixed time intervals used in previous studies (Odean, 1998; Von Beschwitz and Massa 2020; Brettschneider et al., 2021; Ballestra et al., 2024). By allowing investors to adjust between positive and negative portfolio states based on real‑time trading outcomes, our methodology reflects the natural fluctuations in investors’ portfolios, improving the accuracy of our analysis. 

In line with the existing literature, we examine whether taking short exposure influences the disposition effect,  hypothesising that inverse ETF traders, who tend to be more informed and capable, will exhibit a weaker disposition effect compared to typical retail investors (Kelley and Tetlock, 2017; Chen et al., 2007; Da Costa et al., 2013; Von Beschwitz and Massa, 2020). Moreover, we consider that the disposition effect is also modulated by the risk that each trader is willing to accept (Kahneman and Tversky, 1979). 

A further contribution concerns the measurement of risk in leveraged portfolios.  While the prevailing literature typically relies on portfolio standard deviation (e.g., Ballestra et al., 2024; Sakaguchi et al., 2019; Grinblatt and Han, 2005; Barberis, Huang, and Santos, 2001), such measures often conflate market-wide volatility with asset-specific components. By focusing on leveraged ETFs that track the same FTSE MIB index at different leverage ratios, both long and short,
we isolate variation in the intensity of exposure to a common underlying market factor, thereby obtaining a cleaner proxy for relative exposure.

In line with Prospect Theory, we expect higher systematic risk exposure to amplify the disposition effect, with potentially different impacts for long and short exposures. We posit that volatility functions as a behavioral amplifier, increasing the emotional salience of gains and losses and, hence, the strength of the disposition effect. Specifically, we argue that the emotional friction associated with realizing losses, central to Regret Theory, is exacerbated in volatile environments, particularly for retail investors lacking sophisticated hedging capabilities. We therefore predict a differential impact across position types: the disposition effect is expected to be significantly more pronounced in long ETF positions relative to short ETFs, with this asymmetry widening under conditions of high volatility. Moreover, we hypothesize that investors taking short exposure, being more sophisticated, exhibit a weaker disposition effect under narrow framing; however, we examine whether this behavior is moderated by portfolio size, potentially affecting both its magnitude and sign.
Thus, our analysis jointly tests how short exposure, systematic risk, and portfolio context interact in driving realization patterns.

Specifically, to test our hypotheses, we adopt the integrated framing approach proposed by Ballestra et al. (2024). This framework is particularly suited to our analysis because it allows us to measure the disposition effect at the aggregate portfolio level while simultaneously accounting for volatility driven by the common underlying asset of the ETFs and explicitly distinguishing between long and short exposures. In doing so, it overcomes key limitations of earlier approaches. Traditional narrow framing (Odean, 1998) evaluates trades in isolation, ignores the overall portfolio balance, relies on end-of-day snapshots, and is not designed to accommodate inverse positions or short exposure explicitly. Wide framing (Brettschneider et al., 2021) incorporates portfolio context but departs from Odean’s standard measures, reducing comparability with the canonical literature. Our approach bridges this divide by extending Odean’s Count and Total measures to both wide and integrated framing. In addition, we introduce a novel Value metric that captures the percentage return required to realize gains versus losses, allowing us to assess whether investors apply asymmetric realization thresholds. 
To handle the computational burden generated by high‑frequency investor‑level data and the integrated framing metrics, we developed \texttt{dispositionEffect}, an open‑source R package that implements Odean’s measures and our extensions for wide and integrated framing on large datasets (available on CRAN and GitHub).
This framework reconciles stock-level granularity with portfolio-level analysis and provides a unified empirical setting for studying the disposition effect across different framing perspectives.

The remainder of this paper is organized as follows. Section 2 provides a comprehensive literature review that establishes the theoretical foundations of the disposition effect and identifies key gaps in existing research. Section 3 describes our dataset and investor characteristics, highlighting the unique features of our retail investor database and the leveraged ETFs under analysis. Section 4 details our methodology, explaining how we extend Odean's approach to incorporate integrated framing and introducing our novel measures. Section 5 presents our empirical results, analyzing portfolio-driven disposition effects under both narrow and wide framing. It further extends the analysis by adopting an integrated framing approach to examine how short exposure and systematic risk exposure jointly influence the disposition effect. Section 6 concludes by summarizing our key contributions and suggesting directions for future research.


\section{Literature review}
\label{sec: Literature Review}

The disposition effect has traditionally been studied with a focus on long-only investors and conventional determinants. However, recent theoretical and empirical developments have highlighted the importance of exploring this phenomenon through new lenses. Specifically, we identify two critical areas of evolution: investor characteristics, such as the propensity for short exposure and systematic risk, and investor heterogeneity; and the mental accounting framework, ranging from narrow to wide and integrated framing. This review synthesizes existing contributions along these dimensions, highlighting our contribution as a unified empirical framework that examines these factors jointly, particularly in the context of retail investors, who are shown to be most susceptible to the bias.


\subsection*{Short selling and the disposition effect}
The relationship between short selling and the disposition effect has been explored only in a limited number of studies. Von Beschwitz and Massa (2020) show that short sellers display a weaker disposition effect compared to long investors, as they tend to be more sophisticated and better informed. Their analyses, however, rely on aggregated brokerage and securities lending data, which restricts the ability to capture heterogeneity at the individual level.  Kelley and Tetlock (2017) further highlight that retail investors increasingly gain negative market exposure through ETF-based trading strategies, but their study does not explicitly connect this practice to the disposition effect. The practice of short selling faces significant constraints, including costs and availability of borrowable stocks (D'Avolio, 2002), as well as regulatory limitations that can lead to more conservative portfolio management (Massa, 2003). Overall, the literature has primarily considered short selling in terms of informed versus uninformed trading. This focus, often reliant on aggregate data, leaves the specific behavioral dynamics of retail short exposure, and the validity of the disposition effect in this context, largely unexplored.


\subsection*{Systematic risk and the disposition effect}

A closely related strand of the literature studies the role of risk in shaping realization decisions, although it has done so mostly through the lens of idiosyncratic rather than systematic risk. Barberis and Xiong (2012) develop a realization utility model in which the utility from realizing gains is increasing in the idiosyncratic risk of the asset. Their framework implies that stock-specific volatility can amplify the disposition effect by increasing the salience of realized outcomes, but the prediction remains largely theoretical and has not been subjected to a direct empirical test. Along similar lines, Kumar (2009b) shows that behavioral biases, including the disposition effect, are more pronounced for stocks that are harder to value, with uncertainty proxied by characteristics such as idiosyncratic volatility, turnover, and firm age. Likewise, Vasudevan (2023) finds that increases in a stock's volatility during the holding period strengthen the disposition effect, which is interpreted as a response to changes in the perceived riskiness of the individual position. Finally, Sakaguchi et al. (2019) emphasize the importance of portfolio-level volatility for investors' realization choices, but their measure does not disentangle market-wide and stock-specific components of risk.

Taken together, these contributions suggest that the disposition effect is sensitive to risk-related characteristics, but they leave open which dimension of risk is actually driving the result. In particular, the existing evidence mixes conceptually distinct objects, such as idiosyncratic volatility, valuation uncertainty, perceived risk, and aggregate portfolio volatility. As a result, this literature is highly informative about the role of uncertainty in realization behavior, yet it does not identify whether investors respond to stock-specific risk, to common market exposure, or to a combination of the two.

Adopting the stricter perspective of this paper, we view \emph{systematic risk} as variation in exposure to a common market factor rather than as asset-specific uncertainty, subjective risk perceptions, or broad market conditions. On this basis, the literature provides only limited direct evidence. A first group of studies examines the disposition effect across \emph{market states}. Cheng et al. (2013) show that the disposition effect is stronger in bear markets than in bull markets, while Bernard et al. (2021) document that it is countercyclical and significantly stronger during busts than during booms. These papers clearly show that the disposition effect varies over the market cycle, but bull and bear regimes capture joint shifts in beliefs, sentiment, and risk aversion, rather than clean differences in exposure to the same underlying market factor.

A second group of papers examines more broadly the role of uncertainty, volatility, and perceived risk. Kumar (2009b) provides an insightful contribution by showing that behavioral biases, including the disposition effect, are more pronounced for stocks that are harder to value, with uncertainty proxied by characteristics such as idiosyncratic volatility, turnover, and firm age. Vasudevan (2023) offers particularly relevant evidence by showing that increases in a stock’s volatility during the holding period strengthen the disposition effect, interpreting this result as a response to changes in the perceived riskiness of the individual position. Ballestra et al. (2024) complement this line of research by showing that realization decisions depend not only on portfolio performance but also on asset-specific characteristics, including perceived risk. Taken together, these interesting studies provide clear evidence that risk-related factors matter for realization behavior. However, the notion of risk they capture remains tied to valuation uncertainty, changes in individual volatility, or subjective risk perceptions, rather than to a sharp measure of systematic market exposure

Another related strand studies the \emph{aggregate market consequences} of the disposition effect rather than the effect of systematic risk on the disposition effect itself. Grinblatt and Han (2005) show that the disposition effect can have equilibrium pricing implications through capital gains overhang and may help explain momentum. More recently, Yang and Yang (2025) find that changes in the disposition effect predict stock returns beyond standard factor models, while Cui et al. (2025) show that market-wide disposition-related states forecast aggregate volatility. These contributions are very relevant for behavioral asset pricing, but they address a different question, namely how the disposition effect affects prices and volatility at the aggregate level, rather than whether investors exhibit a stronger disposition effect when holding assets with greater systematic risk exposure.

Overall, the literature has made important progress in linking the disposition effect to risk-related variables, market regimes, and asset-pricing consequences. Yet the role of systematic risk remains only partially understood, since existing studies do not isolate differences in risk across securities linked to the same underlying index. The relation between systematic risk and the disposition effect therefore remains an open empirical question, particularly when such securities entail different intensities of exposure.


\subsection*{Investor characteristics}
The strength of the disposition effect is also shaped by investor characteristics. Chen et al. (2007) show that wealthier and more active traders in Asian markets exhibit weaker disposition effects. Da Costa et al. (2013) find that more experienced investors are less prone to this bias. Kumar (2009) demonstrates that individual investor behavior varies significantly based on demographic and psychological factors, with less sophisticated investors showing stronger disposition effects. Von Beschwitz and Massa (2020) confirm that short sellers, who tend to be more sophisticated, display reduced disposition effects relative to typical retail investors. Despite these valuable distinctions, studies have primarily examined investor characteristics via cross-sectional differences in sophistication and experience. To date, no study has jointly analyzed short exposure, systematic risk exposure, and investor heterogeneity within a unified empirical framework. Building on these results, our study deliberately focuses on retail investors. By restricting the analysis to this subgroup, we avoid the heterogeneity that characterizes mixed datasets and concentrate on the type of traders consistently found to be most affected by the disposition effect. This design choice allows us to validate and extend previous findings while ensuring comparability with earlier research on investor sophistication.


\subsection*{Mental accounting framework: Narrow, wide, and integrated framing}
A central methodological issue concerns how the disposition effect is measured. Most studies adopt a narrow framing perspective, pioneered by Odean (1998), which evaluates each trade in isolation. This approach is rooted in the psychological foundations of prospect theory (Kahneman and Tversky, 1979), mental accounting (Thaler, 1980, 1985), regret aversion (Loomes and Sugden, 1982; Fogel and Berry, 2010), and self-control theory (Thaler and Shefrin, 1981). While the narrow framing approach is simple and widely applied, it has been criticized for ignoring portfolio-level considerations. Benartzi and Thaler (2001) introduced the concept of wide framing, highlighting how investors evaluate performance across their entire portfolio. Brettschneider et al. (2021) provide evidence that wide framing amplifies the disposition effect, as investors delay realizing losses to protect perceived overall performance. Ballestra et al. (2024) extend this line of research by proposing an integrated framing approach, in which stock-level and portfolio-level information are jointly considered. Their framework shows that investment decisions are influenced not only by aggregate portfolio performance but also by asset-specific characteristics, such as trading history and perceived volatility. This development highlights the limitations of narrow framing and positions both wide and integrated approaches at the forefront of current research on the disposition effect. \\

Our study contributes to this literature in three main ways. First, we integrate short exposure and systematic risk exposure into a unified empirical framework, an area that remains unexplored. Second, we explicitly focus on retail investors, a subgroup shown by prior literature (Chen et al., 2007; Da Costa et al., 2013; Kumar, 2009; Von Beschwitz and Massa, 2020) to be particularly susceptible to behavioral biases, allowing us to both validate and extend existing findings. Third, we employ an integrated framing methodology that combines narrow and wide perspectives, building on Ballestra et al. (2024) while extending their framework to clearly distinguish between idiosyncratic and market-wide risk and to incorporate short exposure. Unlike previous work, our dataset consists of retail investors trading FTSE MIB-tracking ETFs with both long and short positions, allowing us to examine how these three dimensions, short-selling, systematic risk exposure, and investor heterogeneity, interact within a comprehensive framework.


\section{Data}
\label{sec: Data}

The rise of new technologies and trading platforms has expanded investors’ trading possibilities. Unlike at the end of the last century, investors can now observe real-time price movements throughout the trading day via apps and computers, allowing a deeper analysis of the disposition effect through intraday behavior.

Our dataset, sourced from Directa, an Italian securities brokerage firm, covers approximately 4,600 Italian retail investors from January 2010 to December 2018. The full dataset contains more than 19,000 distinct assets and almost 9 million transactions. Within this broader universe, our main empirical analysis focuses on about 700,000 transactions in eight FTSE MIB-linked leveraged ETFs, with leverage factors of 1x, 2x, 3x, and 7x, including both long and inverse products (-1x, -2x, -3x, -7x). The broader asset universe is used to reconstruct each investor’s portfolio state at the time of ETF trades. To handle the computational burden of these high-frequency investor-level data, we develop the R package \textit{dispositionEffect}, which implements narrow, wide, and integrated framing measures directly from transaction records.

The analysis window was specifically selected to encompass a period with approximately symmetrical movement in the underlying FTSE MIB, ensuring no systematic advantages in being long or short. 
The trading operations are chronologically sorted by execution time and include information on purchase prices and traded quantities. 

The dataset structure displays real-time market prices to assess each position intra-daily allowing us to quantify the net variation in portfolio holdings for individual assets by each client on a daily basis. That way, we can monitor investor behavior within a real-time framework, capturing data related to each transaction even when multiple buy and sell actions occur for the same asset. For intraday market prices, we constructed our own price time series by aggregating prices and timestamps from each client's transactions, creating a market price time series that precisely aligns with the trading times in our dataset.

In particular, our dataset includes all financial transactions an investor made during a specific period. Each transaction contains six key pieces of information: the investor ID, the asset ID, the type of transaction (buy or sell), the quantity traded, the price at which it was traded, and the date and time of the transaction. Furthermore, we can leverage the market prices data, which consists of the computed prices of each traded asset at the time of each transaction.

\begin{table}[H]
    \centering
    \renewcommand{\arraystretch}{1.2} 
    \label{investors-summary}
    \captionsetup{font=footnotesize, labelfont=footnotesize}
    \begin{tabular}{@{}l r@{}}
        \toprule
        \textbf{Metric} & \textbf{Median Value} \\
        \midrule
        Number of transactions & 415 (8 per month) \\
        Number of different assets traded & 32 per investor \\
        Client account time horizon (years) & 2.93 \\
        Holding time horizon per asset (days) & 34.8 \\
        \bottomrule
    \end{tabular}
    \caption{Descriptive Summary of Investors}
\end{table}

The investors in our sample can be classified as active traders, with frequent turnover of positions and relatively short investment horizons. Their behavior is consistent with the profile of active retail traders documented in earlier brokerage studies such as Odean (1998) and Dhar and Zhu (2006). 
In addition, compared to Brettschneider et al. (2021), who use the same dataset as Odean (1998), our data allow us to examine the role of wide framing in a setting that is more focused on intraday transactions and active trading behavior.


\section{Methodology: Disposition effect by asset}
\label{sec: Methodology}

We start our analysis by computing Odean’s (1998) standard measures for realized gains (RG), realized losses (RL), paper gains (PG), and paper losses (PL). Realized gains or losses occur when an investor closes a position in their portfolio, while paper gains or losses include all open positions at the moment of the transaction and all partially closed positions.  Once gains and losses are computed, it is possible to evaluate both the disposition effect of the investor and of each traded asset. Then disposition effect is defined as: 

\begin{displaymath} 
DE = \frac{RG}{RG+PG} - \frac{RL}{RL+PL}
\end{displaymath}

The DE varies between -1 and 1. Positive DE values indicate the presence of a disposition effect, while negative values indicate the presence of an inverse disposition effect. A value of zero indicates that no disposition effect exists.
Further details on the transactions and disposition effect structure are available in the Appendix (Table 8 and 9).

Conventionally, these measures are computed at the investor level by aggregating all transactions into a single pool of gains and losses, irrespective of the specific asset traded (Odean, 1998). This approach implicitly assumes that the bias is invariant across different assets. Our dataset allows us to refine this standard practice by computing the disposition effect for each individual asset traded by the investor. This granularity reveals potential heterogeneity in investor behavior that aggregate measures may mask. To maintain comparability with the existing literature, an overall investor-level disposition effect can then be obtained by aggregating the asset-specific disposition-effect measures.
This asset-level differentiation represents a first methodological advancement and provides the necessary foundation to test whether systematic risk exposure and inverse positions modulate the disposition effect.

In calculating realized and paper measures, Odean (1998) employed two approaches. The Count method treated each trade in an asset as a single observation, regardless of the number of shares traded, and computed gains and losses based on the net difference between end-of-week prices. The total method considered the total number of shares traded, meaning that the realized or paper measure was updated by the exact quantity involved at the end of the week. While we adhere to Odean's methodology, we refine it by updating the measure after each trading operation performed by the investor. Thus gains and losses are computed in real-time rather than being aggregated at weekly intervals reducing aggregation bias in the DE measure. 

We operationalize investors' realized and paper gains and losses using three distinct measures:

\begin{itemize}
\item \textbf{Method Count}: Corresponds to Odean's (1998) standard approach, where gains and losses are computed as simple trade counts.
\item \textbf{Method Total}: Mirrors Odean's alternative methodology, summing the total quantities traded.
\item \textbf{Method Value}: A novel measure where paper and realized gains/losses are expressed in percentage terms. This metric allows us to analyze whether investors require a higher return magnitude before realizing a loss compared to a gain.
\end{itemize}

This framework is particularly suited to examining how systematic risk exposure and short exposure positions modulate the disposition effect, as it allows us to isolate these factors while controlling for investor characteristics and portfolio context.


\subsection*{A portfolio-driven disposition effect by asset via integrated framing}

Having the disposition effect calculated for each asset is the prerequisite for incorporating the portfolio effect into the disposition effect analysis. This allows us to verify whether, as argued by Brettschneider et al. (2021) and Ballestra et al. (2024), the aggregate portfolio balance influences the disposition effect of individual assets. Furthermore, this approach enables us to examine how this relationship might vary across different levels of exposure intensity and position type.

To analyze the portfolio-driven disposition effect, we implement an integrated framing approach (Ballestra et al., 2024) by first applying a filter to determine whether each transaction occurred within a positive or negative portfolio context. This preliminary step ensures that all subsequent computations are categorized accordingly. As a result, each investor's transactions are divided into two groups: those made under a positive portfolio balance and those under a negative portfolio balance.

This classification is based on the cumulative monetary balance of all trading positions after each trade. 
Specifically, for each trading order submitted by an investor, we identify all other positions in different stocks that are open at that point in time and determine whether they are at a gain or a loss. We then sum these positions in monetary terms to conclude whether, at the moment the initial trade occurred, the investor was in a positive or negative portfolio context.
This process is repeated for every investor and is specific to each individual, as the calculation is applied after each trade, only on the portfolio of the investor who made the specific trade.

We are thus able to apply a time frame for investor decisions that is investor-specific. Instead of determining in advance how often to update portfolios (i.e., an arbitrary moment for calculating paper and realized gains/losses as in previous literature by Odean 1998; Brettschneider et al. 2021; and Ballestra et al. 2024), our analysis is based on real investor-specific time frames. This approach was not feasible in the past due to a lack of available data or the computational complexities this type of procedure requires. Thanks to our package \textit{dispositionEffect}, computational problems and the required computing power are overcome.

Additionally, in our methodology, an investor can shift between the positive and negative portfolio groups depending on their trading outcomes at any given time. We evaluate whether the investors' portfolios are in a state of overall gain or loss based on their open trading positions. The rationale behind this is that investors consider their aggregate portfolio balance when making trading decisions. According to prospect theory or regret theory, being in an aggregate gaining situation is expected to make investors more risk-averse, leading them to sell positions that are at a gain more readily. Thus, the probability of selling a gain is higher in the positive portfolio context. Conversely, in line with prospect and regret theory, investors are likely to hold onto their losses, believing they are temporary and waiting for a rebound. In the negative portfolio context, investors are in the risk-seeking domain of the value function, making them more inclined to let their profits run. Therefore, we expect a lower disposition effect in the negative portfolio case.


\subsection*{Testing asymmetry of the disposition effects}

To rigorously test the statistical significance of the disposition effect, we rely on the non-parametric Mann-Whitney U-test. This test is particularly suitable for our analysis for several reasons. Firstly, the Mann-Whitney U-test does not assume a normal distribution of the data. Given the right-skewness observed in our disposition effect distributions, this characteristic is crucial as it allows us to analyze the data without the assumption of normality, which might not hold true for our dataset. Secondly, the test compares the medians of two independent samples. This is particularly useful as the median is a more robust measure of central tendency than the mean. Additionally, the Mann-Whitney U-test is less sensitive to outliers compared to parametric tests like the t-test. This robustness is beneficial for our analysis, as financial data often contains extreme values that can skew results. Moreover, the test can be applied to small sample sizes and is effective even when the sample sizes of the groups being compared are unequal.

By employing this statistical approach, we ensure that our analysis is statistically valid and robust, addressing potential issues related to normality assumptions and providing reliable results. This allows us to confidently interpret the significance of the disposition effect observed in our sample, particularly when examining asymmetries between long and short positions and across different levels of systematic risk.
Specifically, we test whether the difference between the two medians is statistically significant—that is, significantly different from zero. When the result is significant, we further determine which group exhibits the higher median value by examining the sign of the difference (first minus second): a positive value indicates that the first group has the higher median, while a negative value indicates that the second group does.


\section{Results and discussion}
\label{sec: The wide framing}

\subsection*{Data validation: The narrow framing}

\begin{figure}[ht]
    \centering
    \begin{minipage}{0.45\textwidth}
    \caption{Histogram of the disposition effect (method Count)}
    \includegraphics[width=\textwidth]{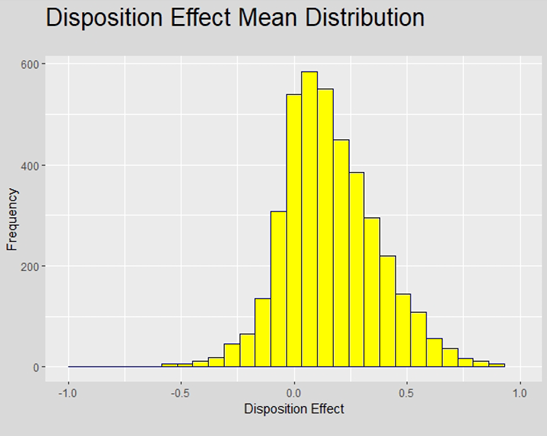}
    \label{fig:hist_count}
    \end{minipage}
    \hfill
\begin{minipage}{0.45\textwidth}
    \centering
    \caption{Histogram of the wide framing disposition effect (method Count)}
    \includegraphics[width=\textwidth]{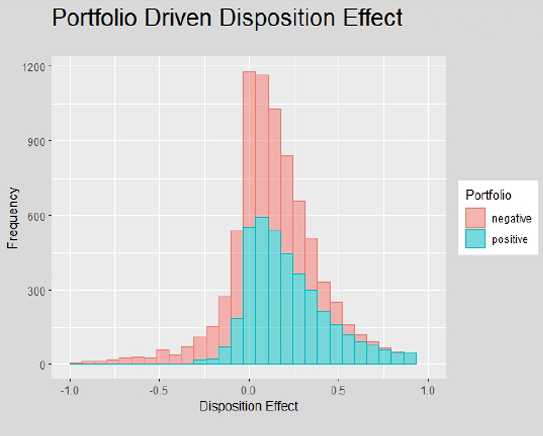}
    \label{fig:hist_portfolio}
\end{minipage}
\end{figure}

As a first step, we validate our database to ensure our data align with previous literature, particularly Odean (1998). For this purpose, we calculated the aggregated disposition effect without differentiating by asset. Figure \ref{fig:hist_count} shows that the distribution of the disposition effect in our sample supports the hypothesis of the disposition effect, exhibiting a mean greater than zero.

We note that our clients predominantly exhibit a positive disposition effect, ranging between 0 and 0.5. Referring to Odean (1998), which documents a disposition effect of 0.21, our sample corroborates this finding with an average disposition effect of 0.20. On average, our sample demonstrates behavior consistent with the disposition effect across all transacted stocks. A key observation from these distributions is the right-skewness of the disposition effect, indicating a deviation from a Gaussian distribution.


\subsection*{Wide framing (a first inspection on the portfolio-driven DE hypothesis)}

Expanding to the wide framing, we observe that investors with losing positions that exceed the monetary value of their winning positions, thus having a negative aggregate portfolio, exhibit a lower tendency to hold onto losing stocks for a longer period. 
Consequently, the distribution for the negative-portfolio group is more concentrated around lower disposition-effect values.
This indicates that the negative portfolio exhibits a lower aggregate disposition effect. The negative portfolio group seems to be linked to a smaller disposition effect, in some cases, even an inverse disposition effect. This result is shown in the histogram (Figure \ref{fig:hist_portfolio}), where the negative portfolio group seems to have the majority of the observations distributed in the 0.0 - 0.5 portion of the graph, but also with numerous observations in the inverse disposition effect.

From the previous distribution, it is evident that our disposition effect remains right-skewed for both alternative portfolio scenarios. Consequently, we rely more confidently on the Mann-Whitney U-test to analyze the data. 

The observed lower disposition effect in negative portfolios can be theoretically interpreted through regret and prospect theories. In our hypothesis, the aggregate portfolio loss could induce investors to view their individual positions as part of a broader, generalized loss. Consequently, they might be more inclined to sell losing stocks to mitigate and reduce the overall loss. As found in Brettschneider et al. (2021), when investors sell stocks at a gain within a negative portfolio, they also tend to sell stocks at a loss. This behavior increases the likelihood of selling losses in the negative portfolio, thereby reducing the paper losses. Consequently, the final effect is a reduction in the aggregate disposition effect. Whether this behavior is mainly induced by the curvature of the value function or the degree of regret aversion is beyond the scope of this paper and is left to future research on the topic.
In conclusion, our aggregate portfolio preliminary results confirm the findings of Brettschneider et al. (2021) and are consistent with Ballestra et al. (2024), the latter of which relied on the same dataset. Hence, the aggregate portfolio result ensures the reliability of our methodology, which allows for the implementation of the wide framing dimension.


\subsection{Integrated framing: The effects of short exposure and systematic risk exposure}
\label{sec: Integrated Framing}

\subsubsection{How systematic risk exposure affects the disposition effect}

Following Ballestra et al. (2024), we adopt an integrated framing approach that jointly considers asset-level and portfolio-level information. We extend their framework to leveraged ETFs and explicitly account for position type by distinguishing between long and inverse exposures.

At the same time, Ballestra et al. (2024) do not distinguish between asset-specific and market-wide volatility, so their aggregate volatility measure cannot determine whether the observed effects reflect idiosyncratic risk or broader market conditions. We address this issue by focusing on ETFs linked to the same FTSE MIB index but characterized by different leverage ratios (1x, 2x, 3x, and 7x for long exposures; -1x, -2x, -3x, and -7x for inverse exposures). This setting allows us to keep the underlying market factor fixed while varying the intensity of exposure to it, thereby providing a sharper way to examine how systematic risk shapes investor behavior and the disposition effect.

Our hypothesis, grounded in prospect theory, posits that higher systematic risk exposure increases the magnitude of gains and losses, thereby amplifying the disposition effect due to the asymmetric curvature of the value function. The concave shape for gains and convex shape for losses implies that investors experience the pain of losses more intensely than the pleasure of equivalent gains. When volatility increases, larger price fluctuations should intensify emotional responses, making investors more likely to realize gains to avoid potential losses and more likely to hold onto losses in anticipation of a rebound.

\begin{table}[H]
\centering
{
\def\sym#1{\ifmmode^{#1}\else\(^{#1}\)\fi}
\begin{threeparttable}
\caption{Variations in disposition effect due to volatility: Long case}
\label{tab:long_volatility}
\begin{tabular}{lD{.}{.}{-1}D{.}{.}{-1}D{.}{.}{-1}}
\hline \hline \\[-1.8ex]
\multicolumn{4}{c}{\textbf{Negative Portfolio}} \\ 
\hline \\[-1.8ex]
\textbf{Long ETF} & \textbf{Count} & \textbf{Total} & \textbf{Value} \\ 
\hline
1x = 2x & -0.050\sym{***} & -0.056\sym{***} & -0.034\sym{***} \\ 
1x = 3x & -0.126\sym{***} & -0.165\sym{***} & -0.130\sym{***} \\ 
1x = 7x & -0.147\sym{***} & -0.139\sym{***} & -0.101\sym{***} \\ 
2x = 3x & -0.076\sym{**} & -0.108\sym{***} & -0.096\sym{**} \\ 
2x = 7x & -0.097\sym{***} & -0.083\sym{***} & -0.068\sym{**} \\ 
3x = 7x & -0.021 & -0.025 & -0.028 \\ 
\hline \hline \\[-1.8ex]
\multicolumn{4}{c}{\textbf{Positive Portfolio}} \\ 
\hline \\[-1.8ex]
\textbf{Long ETF} & \textbf{Count} & \textbf{Total} & \textbf{Value} \\ 
\hline
1x = 2x & -0.138\sym{***} & -0.145\sym{***} & -0.147\sym{***} \\ 
1x = 3x & -0.271\sym{***} & -0.271\sym{***} & -0.322\sym{***} \\ 
1x = 7x & -0.306\sym{***} & -0.282\sym{***} & -0.327\sym{***} \\ 
2x = 3x & -0.133\sym{***} & -0.126\sym{***} & -0.174\sym{***} \\ 
2x = 7x & -0.168\sym{***} & -0.137\sym{***} & -0.179\sym{***} \\ 
3x = 7x & -0.035 & -0.011 & -0.005 \\ 
\hline \hline
\multicolumn{4}{l}{\footnotesize  \sym{*} \(p<0.1\), \sym{**} \(p<0.05\), \sym{***} \(p<0.01\)}\\
\end{tabular} 
\end{threeparttable}
}
\end{table}

\begin{table}[H]
\centering
\def\sym#1{\ifmmode^{#1}\else\(^{#1}\)\fi}
\begin{threeparttable}
\caption{Variations in disposition effect due to volatility: Short case}
\label{tab:short_volatility}
\begin{tabular}{lD{.}{.}{-1}D{.}{.}{-1}D{.}{.}{-1}}
\hline \hline \\[-1.8ex]
\multicolumn{4}{c}{\textbf{Negative Portfolio}} \\ 
\hline \\[-1.8ex]
\textbf{Short ETF} & \textbf{Count} & \textbf{Total} & \textbf{Value} \\ 
\hline
-1x = -2x & -0.036\sym{**} & -0.033\sym{***} & -0.023\sym{***} \\ 
-1x = -3x & -0.084\sym{***} & -0.092\sym{***} & -0.084\sym{***} \\ 
-1x = -7x & -0.175\sym{***} & -0.171\sym{***} & -0.140\sym{***} \\ 
-2x = -3x & -0.048 & -0.059 & -0.062 \\ 
-2x = -7x & -0.139\sym{***} & -0.138\sym{***} & -0.117\sym{***} \\ 
-3x = -7x & -0.091\sym{*} & -0.080\sym{*} & -0.055 \\ 
\hline \hline \\[-1.8ex]
\multicolumn{4}{c}{\textbf{Positive Portfolio}} \\ 
\hline \\[-1.8ex]
\textbf{Short ETF} & \textbf{Count} & \textbf{Total} & \textbf{Value} \\ 
\hline
-1x = -2x & -0.171\sym{***} & -0.174\sym{***} & -0.175\sym{***} \\ 
-1x = -3x & -0.199\sym{***} & -0.190\sym{***} & -0.226\sym{***} \\ 
-1x = -7x & -0.333\sym{***} & -0.225\sym{***} & -0.312\sym{***} \\ 
-2x = -3x & -0.027 & -0.016 & -0.050 \\ 
-2x = -7x & -0.162\sym{***} & -0.051\sym{***} & -0.137\sym{***} \\ 
-3x = -7x & -0.134\sym{*} & -0.035 & -0.087 \\ 
\hline \hline
\multicolumn{4}{l}{\footnotesize  \sym{*} \(p<0.1\), \sym{**} \(p<0.05\), \sym{***} \(p<0.01\)}\\
\end{tabular}
\end{threeparttable}
\end{table}

Our results reveal several key patterns. First, we observe a lower disposition effect in low-leveraged ETFs, with an increase in magnitude corresponding to higher volatility. However, for the 3x vs. 7x comparison, we reject the null hypothesis only for short ETFs, and only at the 90\% confidence interval. This suggests that beyond a certain threshold of volatility, its marginal impact on the disposition effect diminishes. This diminishing sensitivity aligns with the curvature of the prospect theory value function, which posits that the emotional impact of value changes decreases as we move further from the reference point. For extreme variations like those between 3x and 7x ETFs, investor perception may be nearly identical despite the larger volatility exposure of the 7x ETF.

From a behavioral perspective, our analysis supports prospect theory's prediction that investors exhibit risk aversion for gains and risk-seeking behavior for losses. We find that higher asset volatility amplifies the disposition effect, with investors more inclined to realize gains when volatility increases, driven by their aversion to potential losses. The differences observed between long and inverse positions can be attributed to differences in risk perception. Taking short exposure is typically viewed as more complex and risky, which may increase investors’ sensitivity to losses and lead to a more pronounced disposition effect (Kelley and Tetlock, 2017).

A key contribution of our study is the focus on isolating systematic (common factor) risk through leveraged ETFs tracking the same index. By doing so, we address a critical limitation in Ballestra et al. (2024), where market-wide volatility was not differentiated from asset-specific volatility. Our findings provide stronger evidence that both portfolio context and volatility are crucial factors influencing the disposition effect. The integrated framing approach, combining narrow and wide framing dimensions, offers a more comprehensive analysis of investor behavior.

Future research should build on this integrated approach to offer a fuller understanding of the disposition effect. Accounting for both systematic risk exposure and the portfolio context will help refine the model and provide more nuanced insights into investor decision-making, particularly regarding how investor characteristics (such as sophistication) modulate these effects.

\subsubsection{Integrated framing: The effects of short exposure on disposition effect}
\label{sec:Integrated Framing: Effects on short-selling}

In this section, we investigate the presence of the disposition effect among our ETFs and examine whether its magnitude differs between long and inverse positions. Our analysis extends previous work by Von Beschwitz and Massa (2020), who pioneered studying the disposition effect from the perspective of short sellers using weekly data from securities lending companies. While their work provided initial insights, several limitations constrained their findings. First, their dataset structure did not allow for detailed examination of how short sellers operate in financial markets, lacking information on individual investors' portfolios and operations. They relied on estimated purchase prices due to the absence of investor-specific trading data, limiting their analysis to weekly variations in the amount of stocks lent and missing intraday transaction details. Second, their methodology could not account for how investor characteristics, particularly the distinction between professional and retail traders, modulate the disposition effect in short exposure settings.

Our approach addresses these limitations by exploiting investor-level intraday data, which allows us to reconstruct portfolio dynamics in real time rather than relying on aggregated or fixed-interval observations. This granularity enables us to analyze realization decisions conditional on the investor's evolving portfolio state and to examine long and inverse positions within a unified empirical framework. Moreover, by focusing on leveraged ETFs tied to the same FTSE MIB index, we isolate variation in the intensity of exposure to a common market factor, thereby obtaining a sharper measure of systematic risk exposure. This design allows us to compare disposition-effect patterns across position types while abstracting from differences in the underlying asset.

We first tested for differences in the disposition effect under narrow framing. As shown in Table \ref{tab:long_short_narrow}, all ETF coefficients are statistically significant and negative, with consistency across Count-, Total-, and Value-based measures. This marks the first empirical evidence of a disposition effect asymmetry between long and inverse positions. The divergence is notable given that long/short ETF pairs track identical underlying assets. Structural symmetry suggests behavioral biases manifest differently by position direction. Investors taking short exposure appear less prone to the disposition effect, aligning with findings that sophistication reduces disposition effects (Von Beschwitz and Massa, 2020; Kelley and Tetlock 2017; Chen et al., 2007). Under narrow framing, investors likely treat the short ETFs with heightened vigilance; anticipating potential regret (Loomes and Sugden, 1982), they are quicker to cut losses on these speculative positions, thereby dampening the disposition effect.

\begin{table}[H]
\centering
\def\sym#1{\ifmmode^{#1}\else\(^{#1}\)\fi}
\begin{threeparttable}
\caption{Disposition effect differences between long and inverse ETFs}
\label{tab:long_short_narrow}
\begin{tabular}{lD{.}{.}{-1}D{.}{.}{-1}D{.}{.}{-1}}
\hline \hline \\[-1.8ex]
\textbf{ETF} & \textbf{Count} & \textbf{Total} & \textbf{Value} \\ 
\hline \\[-1.8ex]
-1x = 1x & -0.022\sym{**} & -0.025\sym{**} & -0.023\sym{**} \\ 
-2x = 2x & -0.021\sym{**} & -0.021\sym{*} & -0.001\sym{*} \\ 
-3x = 3x & -0.084\sym{**} & -0.087\sym{**} & -0.115\sym{*} \\ 
-7x = 7x & -0.019\sym{*} & -0.045\sym{***} & -0.035\sym{**} \\ 
\hline \hline
\multicolumn{4}{l}{\footnotesize  \sym{*} \(p<0.1\), \sym{**} \(p<0.05\), \sym{***} \(p<0.01\)}\\
\end{tabular}
\end{threeparttable}
\end{table}

However, the narrow framing approach, while useful, has been criticized for its failure to account for portfolio context. To overcome this limitation, we expand our analysis to the integrated framing context by adding the aggregate portfolio dimension to the previous narrow control. Table \ref{tab:long_short_wide} reveals that differences in the disposition effect between long and inverse ETFs are significant in positive portfolios but mixed in negative portfolios. Additionally, in all statistically significant cases, the disposition effect is stronger for inverse ETFs, inverting the narrow-framing findings.

\begin{table}[H]
\centering
{\def\sym#1{\ifmmode^{#1}\else\(^{#1}\)\fi}
\begin{threeparttable}
\caption{Differences in long/short disposition effect: Integrated framing}
\label{tab:long_short_wide}
\begin{tabular}{lD{.}{.}{-1}@{\hskip 0.5cm}D{.}{.}{-1}@{\hskip 0.5cm}D{.}{.}{-1}}
\hline \hline
\textbf{Positive Portfolio} & \textbf{Count} & \textbf{Total} & \textbf{Value} \\ 
\hline \\[-1.8ex]
1x = -1x & -0.029\sym{***} & -0.025\sym{***} & -0.028\sym{***} \\ 
2x = -2x & 0.005 & 0.003 & 0.000 \\ 
3x = -3x & -0.101\sym{*} & -0.107\sym{*} & -0.124\sym{*} \\ 
7x = -7x & -0.002 & -0.083\sym{**} & -0.043 \\ 
\hline \hline
\textbf{Negative Portfolio} & \textbf{Count} & \textbf{Total} & \textbf{Value} \\ 
\hline \\[-1.8ex]
1x = -1x & 0.000 & 0.000 & 0.000 \\ 
2x = -2x & -0.014\sym{**} & -0.024\sym{**} & -0.011\sym{*} \\ 
3x = -3x & -0.042 & -0.073\sym{*} & -0.046 \\ 
7x = -7x & -0.028 & -0.032 & -0.038 \\ 
\hline \hline
\multicolumn{4}{l}{\footnotesize  \sym{*} \(p<0.1\), \sym{**} \(p<0.05\), \sym{***} \(p<0.01\)}\\
\end{tabular}
\end{threeparttable}
}
\end{table}

The divergence between narrow and wide framing results can be reconciled by understanding how the portfolio context alters risk perceptions. Under narrow framing, investors focus solely on the individual position. Recognizing the speculative nature of short ETFs, they manage these positions with heightened vigilance. The fear of losses prompts them to cut losses quickly. This disciplined behavior results in a weaker disposition effect (Von Beschwitz and Massa, 2020; Kelley and Tetlock 2017; Chen et al., 2007).

However, under integrated framing with a positive portfolio, the aggregate gains create a psychological safety net known as the 'house money' effect (Thaler and Johnson, 1990). This buffer reduces the perceived urgency to manage risks on individual speculative trades. Consequently, investors become less likely to close losing short positions, preferring to 'gamble' on a market reversal using the portfolio's surplus gains. This shift from vigilance to risk-seeking behavior inflates the disposition effect for short ETFs, making it stronger than for long ETFs in positive portfolios. Finally, in negative portfolios, consistent with Prospect Theory (Kahneman and Tversky, 1979), the break-even effect triggers uniform risk-seeking behavior across all positions, explaining why the differences between long and inverse ETFs diminish.

We also note that in the case of the negative portfolio, the short ETF might be employed as a hedge against risk when there is concurrent long exposure to FTSE MIB stocks. Thus, the decision to sell the -1x ETF may be influenced by simultaneous decisions to sell FTSE MIB stocks. In such cases, the -1x ETF would likely show profits, while the 1x ETF would show losses. The simultaneous sale of these two positions is consistent with the findings of Brettschneider et al. (2021), which suggest that in a negative portfolio, the probability of selling a losing position increases when a gaining position is sold.

To further investigate how portfolio context affects the disposition effect within each type of ETF, we compare the disposition effect of each ETF in positive versus negative portfolios. As shown in Table \ref{tab:etf_portfolio}, ETFs exhibit a stronger disposition effect in positively performing portfolios, with investors retaining losing positions more frequently in these scenarios. This pattern persists across both long and inverse ETF groups.

\begin{table}[H]
\centering
{\def\sym#1{\ifmmode^{#1}\else\(^{#1}\)\fi}
\begin{threeparttable}
\caption{Differences in ETFs' groups disposition effect: Negative vs Positive portfolio}
\label{tab:etf_portfolio}
\begin{tabular}{lD{.}{.}{-1}@{\hskip 0.5cm}D{.}{.}{-1}@{\hskip 0.5cm}D{.}{.}{-1}}
\hline \hline
 & \textbf{Count} & \textbf{Total} & \textbf{Value} \\ 
\hline \\[-1.8ex]
\multicolumn{4}{l}{\textbf{Short ETFs tested in Negative vs Positive Portfolio}} \\ 
\hline \\[-1.8ex] 
-1x = -1x & 0.000 & 0.000 & 0.000 \\ 
-2x = -2x & -0.136\sym{***} & -0.141\sym{***} & -0.153\sym{***} \\ 
-3x = -3x & -0.115\sym{***} & -0.098\sym{***} & -0.141\sym{***} \\ 
-7x = -7x & -0.159\sym{***} & -0.054\sym{***} & -0.173\sym{***} \\ 
\hline \hline
\multicolumn{4}{l}{\textbf{Long ETFs tested in Negative vs Positive Portfolio}} \\ 
\hline \\[-1.8ex]
1x = 1x & -0.029\sym{***} & -0.025\sym{***} & -0.028\sym{***} \\ 
2x = 2x & -0.117\sym{***} & -0.114\sym{***} & -0.142\sym{***} \\ 
3x = 3x & -0.174\sym{***} & -0.132\sym{***} & -0.220\sym{***} \\ 
7x = 7x & -0.188\sym{***} & -0.168\sym{***} & -0.253\sym{***} \\ 
\hline \hline
\multicolumn{4}{l}{\footnotesize  \sym{*} \(p<0.1\), \sym{**} \(p<0.05\), \sym{***} \(p<0.01\)}\\
\end{tabular}
\end{threeparttable}
}
\end{table}

The stronger disposition effect in positive portfolios corroborates the 'house money' mechanism. When the aggregate portfolio is positive, investors engage in mental accounting (Thaler, 1985), treating losses on individual positions as temporary setbacks that can be absorbed by the overall gains. This fosters an overconfidence in eventual recovery, leading investors to retain losing positions. Conversely, in negative portfolios, the distress of aggregate losses forces a more uniform 'hold-and-hope' strategy, reducing the asymmetry between instrument types. This aligns with Grinblatt and Han (2005) and Kumar (2009), who identify overconfidence and risk aversion as key drivers of the disposition effect.

Our findings highlight the importance of moving beyond narrow framing to understand how investor characteristics and portfolio context modulate the disposition effect. The consistency across long and inverse ETFs underscores the necessity of integrated framing in behavioral analyses. Future work should combine narrow (position-level) and wide (portfolio-level) framing to better model investor decisions, particularly regarding how investor sophistication (retail vs. professional) interacts with position type and portfolio context to influence trading behavior.


\section{Contributions and final remarks}
\label{sec: Conclusion}

This study advances the disposition effect literature by integrating asset volatility, short exposure, and systematic risk exposure into a unified framework that bridges narrow, wide, and integrated framing. We bridge existing gaps in the literature by combining investor-level intraday data, short exposure activity, and portfolio framing within a cohesive empirical setting. By operating within a real-time framework, our approach more accurately captures actual investor behavior and the dynamic nature of decision-making. Consequently, our contribution is both methodological and empirical, offering new insights into investor psychology while providing tools for future research.
Taken together, our findings indicate that the disposition effect is not an asset-specific anomaly, but a core feature of investor behavior that is critically shaped by portfolio context, position type, and exposure to systematic risk.


\subsection*{Methodological contribution}

We extend Odean's (1998) classical measures (Count, Total) to integrated framing and introduce a novel Value dimension, which quantifies the returns or losses required to realize gains. This methodological innovation allows researchers to capture not just whether investors realize gains and losses, but also the significance of the magnitude of returns they require to do so.

Additionally, we introduce \texttt{dispositionEffect}, the first R package dedicated to computing the disposition effect, now freely available on CRAN. This tool resolves the critical computational limitations that previously constrained the scope of empirical research, ensuring a standardized replication of Odean's (1998) methodology while extending it to both narrow (single-asset) and integrated (multi-asset) framing dimensions. By addressing inconsistencies in gain/loss identification and enabling rigorous controls for portfolio-wide outcomes, the package eliminates the reproducibility barriers that hindered prior studies. By providing this open-source tool, we enable future researchers to explore complex behavioral dynamics without facing the computational hurdles that limited previous work.

Moreover, our findings align with established behavioral frameworks, such as Prospect Theory (Kahneman and Tversky, 1979) and Regret Theory (Loomes and Sugden, 1982), while effectively incorporating the dynamics of the 'house money' effect (Thaler and Johnson, 1990) to explain investor decision-making under varying portfolio conditions.

While our results show that Prospect Theory and Regret Aversion both likely contribute to the disposition effect, we do not adjudicate their relative importance. Our findings suggest that these mechanisms coexist in complex portfolios, requiring further empirical work to be fully disentangled. In this regard, the \texttt{dispositionEffect} package offers the methodological support needed to isolate these drivers. By using this tool, researchers can design more granular tests to provide the empirical evidence necessary for refining theoretical models of Regret Theory. Ultimately, the package facilitates the next steps in behavioral research, helping to better define how specific psychological biases influence investor decisions.
In particular, our measures and software infrastructure make it possible to test how the disposition effect varies across portfolio states, levels of systematic risk, and long versus short exposures within a unified empirical design.


\subsection*{Empirical contribution}

Our findings confirm the persistence of the disposition effect across framing paradigms, consistent with foundational work by Odean (1998) and recent extensions by Brettschneider et al. (2021) and Ballestra et al. (2024). Moreover, we extend these insights in several novel directions by examining how short exposure and systematic risk exposure modulate the disposition effect. All three metrics employed (Count, Total, Value) show robust significance, confirming the disposition effect's persistence even when controlling for portfolio-wide outcomes.
This evidence validates the use of integrated framing in high-frequency data and shows that standard disposition measures can be meaningfully generalized beyond the traditional narrow view. By exploiting long and short leveraged ETFs tied to the same FTSE MIB index, we obtain a clean empirical setting in which different intensities of systematic risk exposure can be studied while holding the underlying market factor constant.
Using a large sample of retail investors, we provide evidence of the disposition effect in inverse ETFs linked to the FTSE MIB index, indicating that retail investors exhibit similar biases in short and long exposures.
Notably, our focus on retail trading in inverse ETFs, a previously unstudied context, reveals that disposition biases persist even in leveraged instruments typically associated with informed traders, interestingly with lower magnitude. This confirms that investor sophistication, rather than instrument complexity alone, is the critical factor in mitigating behavioral anomalies (Chen et al., 2007; Von Beschwitz and Massa, 2020).

Moreover, our analysis uncovers a critical portfolio-state asymmetry that challenges the symmetry of long/short decision-making. In the narrow framing context, investors taking short exposure display a more restrained disposition effect, consistent with Regret Theory (Loomes and Sugden, 1982), as the inherent risks associated with inverse leveraged positions induce greater vigilance to avoid accelerating losses. However, under integrated framing, this pattern inverts: investors taking short exposure exhibit a stronger disposition effect in positive portfolios. We attribute this inversion to the 'house money' effect (Thaler and Johnson, 1990), where aggregate gains create a psychological safety net that reduces the urgency to close losing short positions. Conversely, in negative portfolios, the break-even effect of the Prospect Theory (Kahneman and Tversky, 1979), triggers uniform risk-seeking behavior across all positions, explaining why differences between long and inverse ETFs diminish. These results underscore the role of mental accounting in modulating behavioral anomalies (Thaler, 1985), showing that the portfolio context dictates whether investors act with vigilance or risk-seeking speculation.
In this sense, our evidence only partially confirms our initial expectation of a consistently weaker disposition effect among short-exposed investors, revealing instead that the asymmetry between long and short positions is itself state-dependent.
Finally, we show that systematic risk exposure is associated with stronger disposition effect in integrated framing. Our results align with dual theoretical drivers: Prospect Theory (reference-dependent risk preferences) and Regret Aversion (anticipated remorse over realized losses).
By exploiting ETFs with different leverage ratios tied to the same index, we document that higher exposure to the common market factor is associated with stronger realization asymmetries, supporting the view that volatility acts as a behavioral amplifier rather than a purely statistical characteristic of returns.


\subsection*{Limitations and future research}

Despite our contributions, a limitation of our study is the exclusive focus on retail investors, which enhances the behavioral interpretability of our results but may limit their generalizability to professional or institutional traders, who typically adopt different risk‑management strategies. Future studies could compare how framing and systematic risk jointly affect realization behavior for retail and institutional investors within similar leveraged environments.

Additionally, our analysis of short-selling, while innovative, is constrained to ETFs. Future work could examine whether similar patterns emerge in direct short selling of individual stocks, particularly among institutional investors.

In conclusion, our findings show that a narrow, trade‑by‑trade perspective is no longer sufficient to explain how investors realize gains and losses. By jointly accounting for investor characteristics, position type, systematic risk, and portfolio context, we uncover systematic asymmetries in the disposition effect that persist even among retail investors trading complex inverse ETFs and operating in high-risk environments. These results point to a deeply rooted behavioral component in trading decisions and call for renewed attention to how framing and risk interact in shaping investor behavior, highlighting that the disposition effect is not a residual anomaly, but a core and resilient feature of investor decision‑making.
Our unified framework, together with the \texttt{dispositionEffect} package, offers a scalable platform for subsequent empirical work on realization behavior, enabling future studies to refine behavioral asset pricing models and to design policies and interventions aimed at improving investor welfare.

\vspace{20pt}

\noindent \textbf{Data and Code Availability}

\par\noindent
The data underlying this study are proprietary and cannot be publicly shared. However, they are available from the corresponding author upon reasonable request. The code used for the analysis is publicly available on CRAN (\url{https://CRAN.R-project.org/package=dispositionEffect}) and GitHub (\url{https://github.com/marcozanotti/dispositionEffect}).

\newpage


\begin{footnotesize}

\end{footnotesize}

\end{document}